\documentstyle[epsfig]{article}
\begin{document}
\title{Charm production in antiproton-nucleus collisions at the 
$J/\psi$ and the $\psi'$ thresholds}
\author{L.~Gerland$^a$, L.~Frankfurt$^b$, M.~Strikman$^c$\vspace{.5cm}\\ 
{\small $a$ SUBATECH, Laboratoire de Physique Subatomique et des
Technologies Associ\'ees} \\
{\small University of Nantes - IN2P3/CNRS - Ecole des Mines de Nantes} \\
{\small 4 rue Alfred Kastler, F-44072 Nantes, Cedex 03, France,}\vspace{2mm}\\
{\small $b$ School of Physics and Astronomy, Raymond and Beverly Sackler
Faculty}\\
{\small of Exact Science, Tel Aviv University, Ramat Aviv 69978,
Tel Aviv, Israel,}\vspace{2mm}\\
{\small $c$ Pennsylvania State University, University Park, Pennsylvania 
16802, USA}}

\maketitle

\abstract{We discuss the production of charmonium states in
antiproton-nucleus collisions at the $\psi'$ threshold.  It is explained
that measurements in $\bar p A$ collisions will allow to get new
information about the strengths of the inelastic $J/\psi N$ and $\psi'N$
interaction, on the production of $\Lambda_c$ and $\bar{D}$ in
charmonium-nucleon interactions and for the first time about the
nondiagonal transitions $\psi' N\to J/\psi N$. The inelastic
$J/\psi$-nucleon cross section is extracted from the comparison of
hadron-nucleus collisions with hadron-nucleon collisions. We extract the
total $J/\psi$-nucleon cross section from photon-nucleon collisions by
accounting for the color transparency phenomenon within the frame of the
GVDM (Generalized Vector meson Dominance Model). We evaluate within the
GVDM the inelastic $\psi'$-nucleon cross section as well as the cross
section for the nondiagonal transitions. Predictions for the ratio of
$J/\psi$ to $\psi'$ yields in antiproton-nucleus scatterings close to the 
threshold of $\psi'$ production for different nuclear targets are 
presented.}

\section{Introduction}

During the last two decades significant attention was given to the
absorption of charmonium states produced in heavy ion collisions, see
e.g.\ ref.~\cite{vogt} and references therein. An important role in such
evaluation is played by the value of the total and the elastic cross
sections for charmonium-nucleon interactions as well as the amplitude for
the inelastic transition between $J/\psi$ and $\psi'$ states
characterizing the role of color transparency phenomena. The aim of this
paper is to extract these cross sections from photoproduction data
following ref.~\cite{zhalov} and to make predictions for
antiproton-nucleus collisions at the $\psi'$ threshold. We demonstrate
that in these collisions the cross section for the nondiagonal transition
$\psi'+N\to J/\psi+N$ can be measured. We account for the dependence of
the cross sections on energy, and the dependence of the elastic cross
section on the momentum transfer.

The charmonium production at the $\psi'$ threshold is well suited to
measure the genuine charmonium-nucleon cross sections. At higher energies
formation time effects makes the measurement of these cross sections more
difficult~\cite{ger}. These cross sections and the cross section for the
nondiagonal transition $\psi'+N\to J/\psi+N$ are important for the
analysis of charmonium production data at
SPS-energies~\cite{spieles,brat}. At collider energies, i.e.\ at RHIC and
LHC, the formation time effects will become dominant and charmonium states
will be produced only far outside of the nuclei~\cite{ger2}. However,
measurements of the genuine charmonium-nucleon cross sections as well as
the cross section for the nondiagonal transition $\psi'+N\to J/\psi+N$ are
also important at collider energies for the evaluation of the interaction
of charmonium states with the produced secondary particles.

By using the appropriate incident energy in antiproton-nucleus scattering, 
the case when the scattering occurs off the nucleons with small internal 
momenta can be selected. Correspondingly, in this situation
off shell effects in the amplitude should be very small and, hence, there
will be no significant nuclear corrections due to possible modification of
the nucleons in nuclei. In addition, due to the large antiproton-nucleon
cross section, those antiprotons which do not undergo absorption at the 
surface of the nuclear target will lose a significant fraction of their 
energy. Therefore, at energies close to the threshold the charmonium 
production is almost impossible inside of the nuclear target.

We discuss in this paper the production of charm at the $p\bar p\to
J/\psi,\psi'$ thresholds. To avoid difficulties with the specifics of low
energy initial state interaction effects, which are actually included into
the partial width of the $J/\psi \to p\bar p$ decay, like those discussed
in ref.~\cite{Brodsky:1995ds}, we will discuss ratios of cross sections,
in which the factor $\sigma(p\bar p\to J/\psi,\psi')$ is canceled. We
demonstrate that these ratios are well suited to measure the nondiagonal
($\psi'\to\psi$) cross section as well as the inelastic $J/\psi$ and
$\psi'$ cross section. At the same time the momentum of charmonium in the
final state is 5~GeV/c in the rest frame of the nuclear target. Hence, one
cannot probe in this reaction a possible enhancement of the
charmonium-nucleus cross section near threshold of charmonium-nucleus
interactions like that described in ref.~\cite{Brodsky:1989jd}.

The amplitude of the $J/\psi$ photoproduction close to the threshold
$E_{\gamma}\sim 9$~GeV is dominated by the generalized gluon density at
large $x_1-x_2 \sim 0.5$. In such a kinematic, Fermi motion effects may
lead to a significant enhancement. However, similar to the quark
distribution functions one may expect a suppression reflecting medium
modifications of the nucleon structure functions (the analogue of the EMC
effect). In a genuine photoproduction experiment it would be very
difficult to distinguish the EMC type effect from the absorption due to
the final state interaction. However, the combination of a measurement at
the GSI of the $\bar p +A\to J/\psi+X $ reaction and of photoproduction at
12~GeV at the Jefferson-lab will make it possible to measure the
$A$-dependence of the nuclear generalized gluon distributions at large
$x$.

The ratio of the elastic to the total $J/\psi N$ cross section 
has been evaluated long ago in ref.~\cite{Ioffe:kk} in the Vector
meson Dominance Model (VDM), where the energy dependence and the real part
of the forward scattering amplitude were neglected and the $t$ dependence
of the elastic cross section was effectively adjusted to the data on
the soft QCD process of the $\rho$ photoproduction off the nucleon target.
We will show in this paper that all these effects and color
transparency phenomenon should be taken into account.

In the beginning of this paper we examine the energy dependence of the
ratio of the elastic to the total $J/\psi N$ cross section as well as the
influence of the real part of the amplitude and treat the $t$ dependence
of the elastic cross section within a charmonium model. The final result
is that the ratio of the elastic to the total $J/\psi N$ cross section is
still small (approximately $5\%\div 6.5\%$) but by a factor $\approx 2.5$
larger than that given in the previous evaluation.

In the following section the amplitudes of the GVDM, the elastic form
factor and the two-gluon-form factor are described. The amplitudes of the
GVDM are described in more detail in the appendix. In the third section
the semiclassical Glauber model is described and the predictions for the
future GSI experiment are shown. In the fourth section the results of
this paper are summarized. The phenomena considered in this paper are
complementary to the program of antiproton-nucleus scattering
experiments at the GSI range outlined in the recent
review~\cite{Brodsky:2004ah}.

\section{Model description and results}

In ref.~\cite{Ioffe:kk} the elastic and the total $J/\psi$-nucleon cross 
sections were evaluated within the Vector meson Dominance Model
(VDM). In this model, the $J/\psi$ photoproduction amplitude 
$f_{\gamma \psi}$ and the $J/\psi$-nucleon elastic scattering 
amplitude $f_{\psi \psi}$ are related as,
\begin{equation}
f_{\gamma \psi}={e\over f_{\psi}}f_{\psi \psi}\quad .
\label{VDM}
\end{equation}
Here, $e$ is the charge of an electron and $f_{\psi}$ is the 
$J/\psi-\gamma$ 
coupling given by, 
\begin{equation}
\left({e^2\over 4 \pi f_{\psi}}\right)^2={3\over 4\pi}
{\Gamma(V\to e\bar e)\over m_{\psi}}\quad.
\end{equation}
A similar relation like eq.~(\ref{VDM}) can be written also for the 
$\psi'$. From the optical theorem,
\begin{equation}
\sigma_{tot}(J/\psi N)
= {I_{\psi \psi}(t=0)\over 
2p_{cm}E_{cm}}\quad, 
\end{equation}
where $I_{\psi \psi}$ is the imaginary part of $f_{\psi \psi}$ and the 
differential elastic cross section,
\begin{equation}
{d\sigma_{el}/dt}={1\over 64\pi\, p^2_{cm}E^2_{cm}} |f_{\psi \psi}|^2
\label{dsigel}
\end{equation}
follows
\begin{equation}
\sigma_{el}= {\sigma_{tot}^2(J/\psi N)|t_{eff}| (1+\eta^2)\over 
16\pi}\quad.
\label{ioffe}
\end{equation}
$|t_{eff}|$ comes from the integration of eq.~(\ref{dsigel}) over $t$.
$\eta$ is the ratio of the real part to the imaginary part of the 
amplitude of $J/\psi N$ scattering.
The $t$-dependence of the differential cross section is given by the
square of the two-gluon-form factor~\cite{Frankfurt:2002ka}, which is
\begin{equation}
F^2_{2g}(t)={1\over \left(1-{t\over m^2_{2g}}\right)^4}\quad.
\end{equation}
with $m^2_{2g}\approx 1.1\mbox{ GeV}^2$. 

\begin{figure}
    \centering
        \leavevmode\
        \epsfxsize=1.\hsize
        \epsfbox{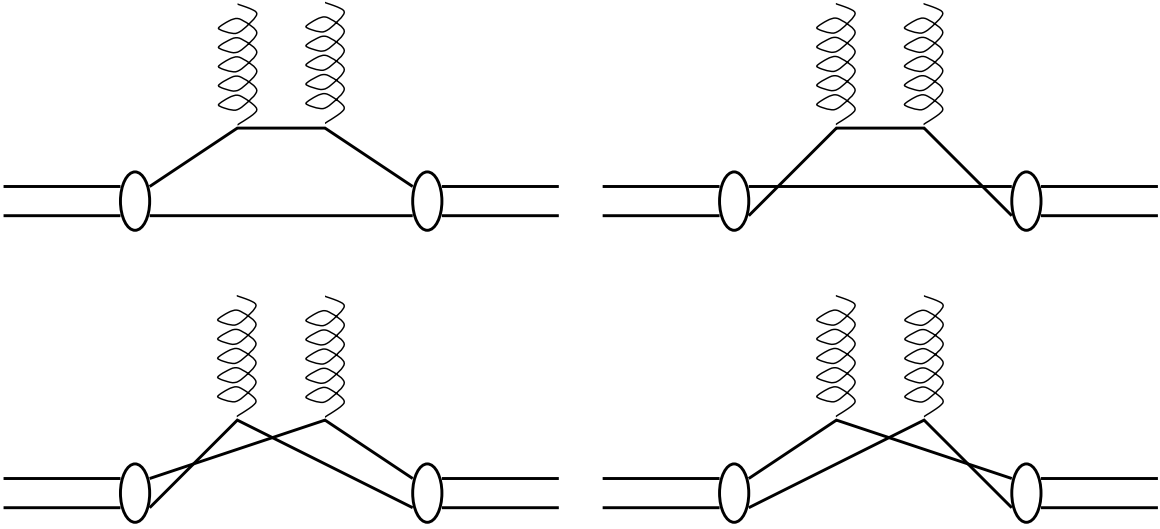}
\caption{The four leading graphs that contribute to the elastic 
form factor of the $J/\psi$}
\label{twog}
\end{figure}

And the two-gluon-form factor
of the $J/\psi$, calculated as the nonrelativistic 
limit of the diagrams shown in fig.~\ref{twog} is 
\begin{equation}
F_{\psi}(t)=\frac {\int \Psi(z,k_t)\Delta(k_t) \Psi(z,k_t-zq_t)
d^2k_t {dz\over z(1-z)}} {\int \Psi(z,k_t)\Delta(k_t)
\Psi(z,k_t) d^2k_t {dz\over z(1-z)}}\quad .
\end{equation}
Here, $\Psi$ is the wave function of the $J/\psi$, $z$ is the fraction of 
the longitudinal momentum of the charmonium state carried by the 
$c$-quark, while $k_t$ is the relative transverse momentum of the $c$-quark 
and the $\bar c$-quark. $\Delta(k_t)$ is two dimensional Laplace operator. 
$q_t$ is the sum of the momenta of the two gluons. This form factor 
unambiguously follows from the analysis of Feynman diagrams for hard 
exclusive processes. By definition it is equal to one at zero momentum
transfer $F_{\psi}(t=0)=1$. To evaluate this form factor we 
use here the nonrelativistic wave functions of ref.~\cite{eich}. 

In the gluon exchange between the charmonium and the target only one
gluon polarization dominates. In QCD evolution only this contribution
contains the large logarithm $\ln (m_c^2)$. Using the QCD Ward identity
one can express the obtained formulae in terms of the exchange by
transversely polarized gluons like in the derivation of the
Weizs\"acker-Williams approximation. In the nonrelativistic approximation
the binding is dominated by a Coulomb potential. The Yang-Mills vertex
between the Coulomb potential and transversely polarized gluons is zero.  
Therefore, only the interaction between the two gluons and the two heavy
quarks of fig.~\ref{twog} have to be taken into account in this
calculation.

\begin{figure}
    \centering
        \leavevmode\
        \epsfxsize=.6\hsize
        \vspace{-1cm}
        \epsfbox{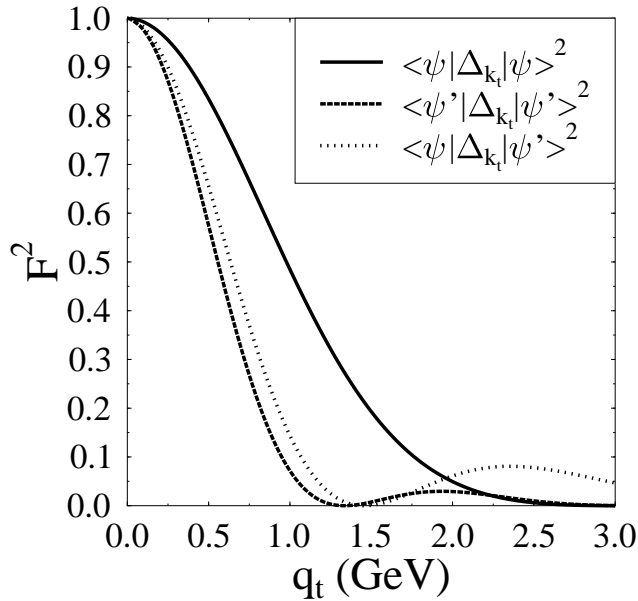}
\caption{The form factors squared of the diagonal ($J/\psi\to 
J/\psi$ and $\psi'\to \psi'$) and the nondiagonal transitions 
($J/\psi\to \psi'$ and $\psi'\to J/\psi$).
}
\label{form}
\end{figure}

\begin{figure}
    \centering
        \leavevmode\
        \epsfxsize=.6\hsize
        \vspace{-1cm}
        \epsfbox{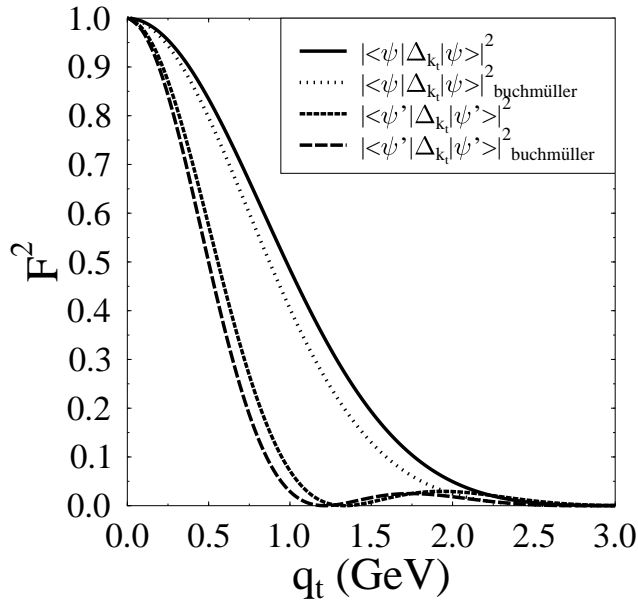}
\caption{The form factors squared of the diagonal ($J/\psi\to
J/\psi$ and $\psi'\to \psi'$) for two different nonrelativistic 
charmonium models.}
\label{form2}
\end{figure}

The nonrelativistic approximation is justified
at small momentum transfer because of the large mass of the $c$-quark. 
Small momentum transfers are the most important domain because the 
two-gluon form factor decreases rather quickly with the momentum transfer. 
The result for the
elastic $J/\psi$ form factor is shown in fig.~\ref{form}. Additionally, 
fig.~\ref{form} depicts the elastic $\psi'$ form factor as well as the 
nondiagonal transition from the $J/\psi$ into the $\psi'$. 
In fig.~\ref{form2} the dependence of the form factor on the charmonium model
is shown. The elastic form factors of the $J/\psi$ and the 
$\psi'$ calculated in two different charmonium models are depicted. The 
charmonium models are from the refs.~\cite{eich,buch}. One can see that 
the dependence on the charmonium model is small in comparison to other 
uncertainties. 
The elastic cross section is then proportional to 
\begin{equation}
|t_{eff}|=\int dt\,F^2_{2g}(t)F^2_{\psi}(t)\quad .
\end{equation}
Integrating the two-gluon-form factor of ($J/\psi\to J/\psi$) over $t$
yields
\begin{equation}
\int dt\,F^2_{2g}(t)= {1\over 3}
m^2_{2g} \left(1-\left({s+m^2_{2g}\over 
m^2_{2g}}\right)^{-3}\right)\approx 0.4\mbox{ GeV}^2\quad.
\label{intform}
\end{equation}
The approximation at the end of this equation is for
sufficiently high energies, where it is possible to neglect the last 
term. Taking into account the elastic form factor of the $J/\psi$ reduces 
this value to $|t_{eff}|=0.3\mbox{ GeV}^2$. 

Eq.~(\ref{intform}) differs from the power law that arises in the limit of
large $t$, i.e.\ in large angle scattering where $-t/s\sim 1/2$. In this
regime, the selection of dominant diagrams follows from the requirement to
obtain the lowest power of $t$.  In the literature this is known as power
counting rules. However, in the processes considered in this paper, this
integral is dominated by $(-t)\cdot r_N^2\sim 1$, where $r_N$ is the
radius of a nucleon. This kinematical region does not overlap with
high-momentum transfers.

Two important phenomena are neglected in the VDM model. One is the color 
transparency phenomenon due to production of $c\bar c$ in configurations 
substantially smaller than the mean $J/\psi$ size. As a result the 
effective cross section $\sigma_{tot}(J/\psi N)$ as extracted from the 
$J/\psi$ photoproduction off a nucleon is much smaller than the genuine 
cross section of the $J/\psi$-N interaction. Another neglected effect is 
the hard contribution to $\sigma_{tot}$ which rapidly increases with 
energy~\cite{FS88}.
Therefore, we use the correspondence between the GVDM and the QCD dipole 
model which leads to the parametrization of cross
section see ref.~\cite{zhalov}
\begin{equation}
\sigma_{tot}(J/\psi N)=3.2\mbox{ mb}\left({s\over s_{0}}\right)^{0.08} +
0.3\mbox{ mb}\left({s\over s_{0}}\right)^{0.2}
\label{psiN}
\end{equation}
with $s_0=39.9\mbox{ GeV}^2$.

It is worth 
noting here that such a parametrization is reasonable only
for the energies where inelastic nondiffractive channels  
(the lowest nondiffractive channel is $\Lambda_c + \bar{D}$) are open, 
that is for $\sqrt{s} > 4.15$~GeV, in 
the rest system of the nucleon this is $\omega > 3.61$~GeV.
The amplitudes within the GVDM are related by
\begin{eqnarray}
f_{\gamma \psi}&=&{e\over f_{\psi}} f_{\psi \psi}+{e\over
f_{\psi'}}f_{\psi' \psi}\cr
f_{\gamma \psi'}&=&{e\over f_{\psi}}f_{\psi \psi'}
+{e\over f_{\psi'}} f_{\psi' \psi'}\quad.
\label{GVDM}
\end{eqnarray}
The amplitudes, $f_{\psi \psi'}$ and $f_{\psi' \psi}$, that appear here
additionally in comparison to the VDM in eq.~(\ref{VDM}) are the
amplitudes for the nondiagonal transitions $J/\psi\to \psi'$ and $\psi'\to
J/\psi$ respectively. The amplitudes following from eq.~(\ref{psiN}) and
eq.~(\ref{GVDM}) are given in appendix~\ref{app}.

\begin{figure}
    \centering
        \leavevmode\
        \epsfxsize=.6\hsize
        \vspace{-1cm}
        \epsfbox{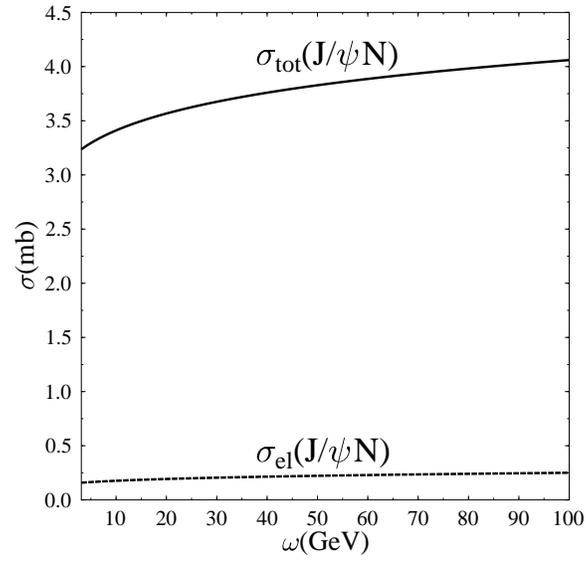}
\caption{The elastic and the total $J/\psi$-nucleon cross section in
dependence of the energy of the $J/\psi$ in the rest frame of the
nucleon.}
\label{absolut}
\end{figure}

\begin{figure}
    \centering
        \leavevmode\
        \epsfxsize=.6\hsize
        \vspace{-1cm}
        \epsfbox{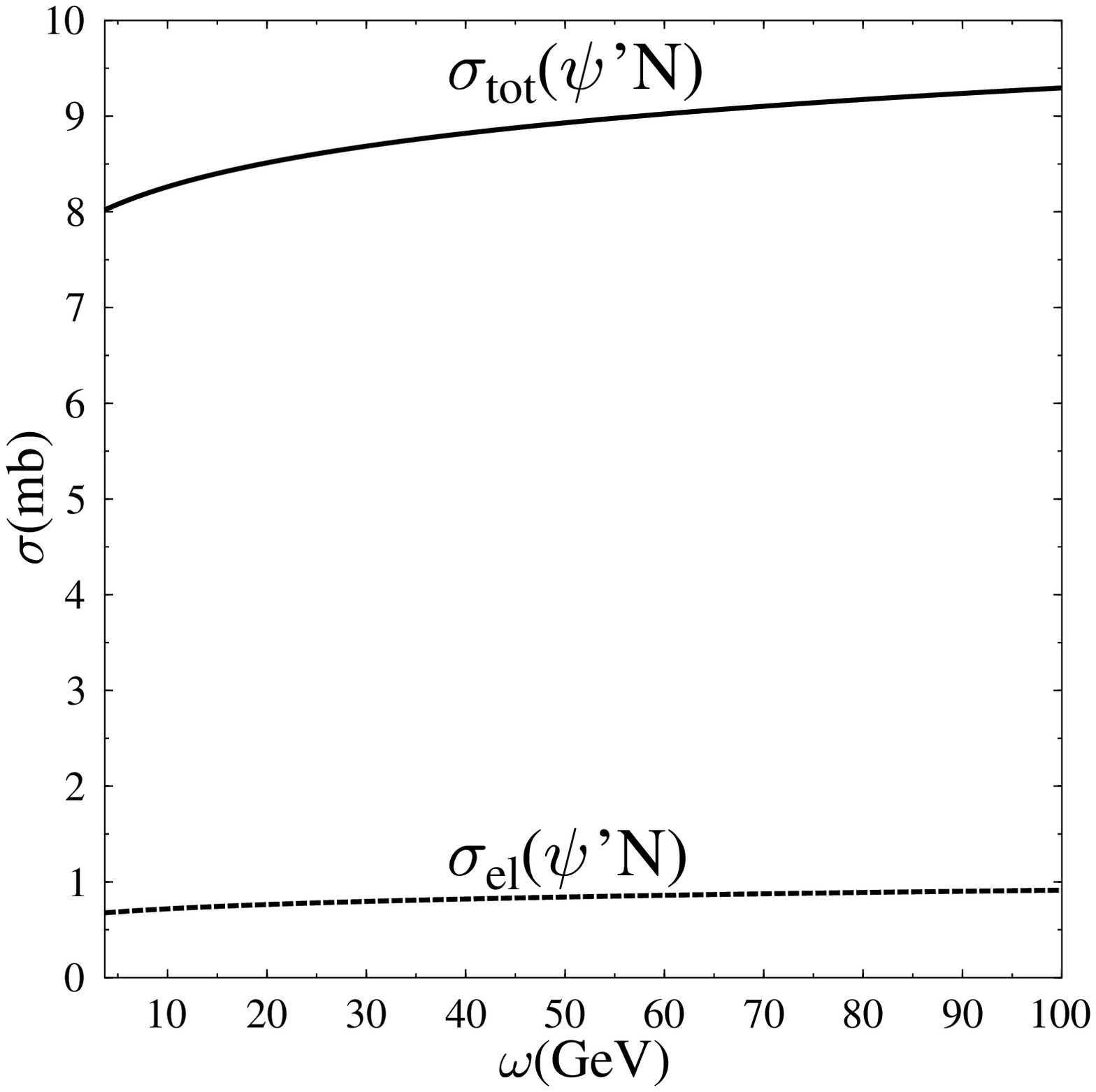}
\caption{The elastic and the total $\psi'$-nucleon cross section in
dependence of the energy of the $\psi'$ in the rest frame of the
nucleon.}
\label{absolutprim}
\end{figure}

\begin{figure}
    \centering
        \leavevmode
        \epsfxsize=.6\hsize
        \vspace{-1cm}
        \epsfbox{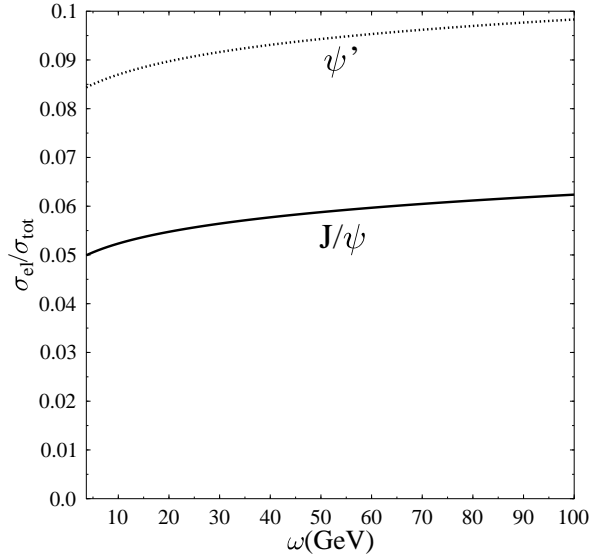}
\caption{The ratio of the elastic charmonium-nucleon cross section to 
the total charmonium-nucleon cross section in dependence of the energy of 
the charmonium in the rest frame of the nucleon.}
\label{ratio}
\end{figure}

The results of eq.~(\ref{ioffe}) and eq.~(\ref{psiN}) (the total and the 
elastic cross section for $J/\psi N$ collisions) are shown in 
fig.~\ref{absolut}. Fig.~\ref{absolutprim} shows the same for $\psi' 
N$ collisions. The ratio of the elastic to the total cross section is 
depicted in fig.~\ref{ratio}.

\begin{figure}
    \centering
        \leavevmode\
        \epsfxsize=.6\hsize
        \vspace{-1cm}
        \epsfbox{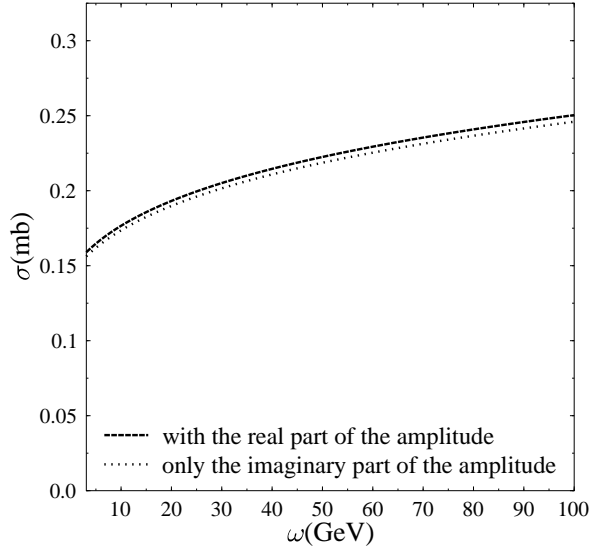}
\caption{The elastic $J/\psi$-nucleon cross section in
dependence of the energy of the $J/\psi$ in the rest frame of the
nucleon is shown with and without the real part of the amplitude.}
\label{absolut2}
\end{figure}

The elastic cross section calculated with and without the real part of the 
amplitude is shown in fig.~\ref{absolut2}. The real part contributes 
approximately $2\%$ to the elastic cross section in the discussed 
energy range. 

\begin{figure}
    \centering
        \leavevmode\
        \epsfxsize=.6\hsize
        \vspace{-1cm}
        \epsfbox{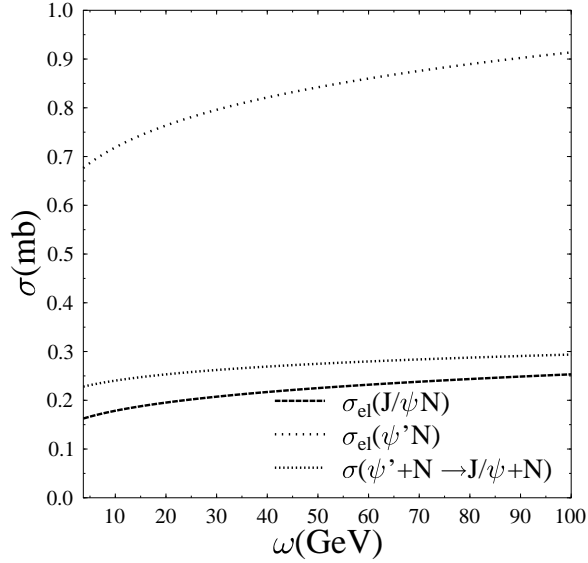}
\caption{The elastic $J/\psi$-nucleon cross section, the elastic 
$\psi'$-nucleon cross section and the nondiagonal cross 
section ($\psi'+N\to J/\psi+N$) in
dependence of the energy of the charmonium in the rest frame of the
nucleon.
}
\label{psipsiprimpcm}
\end{figure}

Fig.~\ref{psipsiprimpcm} shows the energy dependence of the elastic
$J/\psi$-nucleon cross section, the elastic $\psi'$-nucleon cross section
and the nondiagonal cross section ($\psi'+N\to J/\psi+N$). One can see
that the nondiagonal cross section ($\psi'+N\to J/\psi+N$) is comparable
with the elastic $J/\psi$-nucleon cross section.

\section{$\bar p A$ collisions at the $\psi'$ and the $J/\psi$
threshold}

A program of studies of charmonium production in a $\bar p A$ collisions 
at a $\bar p$ accumulator is planned~\cite{gsi}. 
Hence we discuss in this section the production of charmonium states in
the antiproton-nucleus collisions at the $\psi'$ threshold. The direct
production of $J/\psi$'s is suppressed here. However, a $\psi'$ is 
produced and becomes an $J/\psi$ in a further collision with a nucleon
in the nuclear target. Since the produced hidden charm state has a
large momentum relative to the nucleus target the semiclassical 
Glauber-approximation can be used. In our calculation we will neglect 
color transparency effects in the 
initial state for the production of $J/\psi$ and $\psi'$ 
mesons~\cite{Brodsky:1988xz}, since the coherence length for the 
fluctuation of the incoming antiproton into a small configuration is very 
small at the relevant energies practically completely washing out the CT 
effect~\cite{farrar}.

The production of a $J/\psi$ at the threshold in a $\bar p A$ collision
and the subsequent production of a $\psi'$ in a rescattering of the
$J/\psi$, is not well suited for the measurements of the nondiagonal cross
sections. This is because in a $\bar p A$ collision at
$\sqrt{s}=m_{\psi}=3.1$~GeV the $J/\psi$ is produced at rest in the center
of mass system. This means the energy in the center of mass of the
$J/\psi$ and the nucleon is 4.5~GeV. The threshold for the production of a
$\psi'$ in such a collision is $m_{\psi'}+m_{N}=4.626$~GeV. This boundary
is extended when nucleon Fermi motion within the nuclear target is taken
into account (see ref.~\cite{farrar} for the discussion of the role of
Fermi motion effects in the production of charmonium states).

However, this process is strongly suppressed by the phase space. At
the same time, the process $J/\psi +p\to \Lambda_c + \bar D$ 
is likely to dominate the inelastic cross section. Hence the
measurement of the process $\bar p +A \to \Lambda_c + \bar D+X$ in the
vicinity of $s_{\bar p p}=m^2_{J/\psi}$ will allow a direct measurement of
$\sigma_{inel}(J/\psi N)$.

In the semiclassical Glauber-approximation the cross section to produce a 
$\psi'$ in an antiproton-nucleus collision is

\begin{eqnarray}
\sigma\left(\bar p + A\to \psi' \right)&=&2\pi\int db\cdot 
b\,dz_1{n_p\over A}\rho(b,z_1)\sigma\left(\bar p + 
p\to \psi' \right)exp\left(-\int\limits_{-\infty}^{z_1} 
dz \sigma_{\bar p N inel}\rho(b,z)\right)\cr
& &\phantom{xxxxx}
\times exp\left(-\int\limits_{z_1}^{\infty}dz\sigma_{\psi'N 
inel}\rho(b,z)\right)\quad.
\label{antipapsiprim}
\end{eqnarray}

In this formula, $b$ is the impact parameter of the antiproton-nucleus
collision, $n_p$ is the number of protons in the nuclear target, $z_1$ is
the coordinate of the production point of the $\psi'$ in beam direction,
and $\rho$ is the nuclear density. $\sigma\left(\bar p + p\to
\psi'\right)$ is the cross section to produce a $\psi'$ in an
antiproton-proton collision. $\sigma_{\bar p N inel}$ is the inelastic
antiproton-nucleus cross section.  $\sigma_{\psi' N inel}$ is the 
inelastic $\psi'$-nucleon cross section.

All the factors in eq.~(\ref{antipapsiprim}) have a rather direct
interpretation. $exp\left(-\int\limits_{-\infty}^{z_1} dz \sigma_{\bar p N
inel}\rho(b,z)\right)$ gives the probability to find an antiproton at the
coordinates $(b,z_1)$, which accounts for its absorption, and ${n_p\over
A}\rho(b,z_1)\sigma\left(\bar p + p\to \psi'\right)$ is the probability 
to create a $\psi'$ at these coordinates. The factor ${n_p\over A}$ accounts 
for the fact that close to the threshold the antiproton can produce a 
$\psi'$ only in an annihilation with a proton but not with a neutron. The 
term $exp\left(-\int\limits_{z_1}^{\infty}dz\sigma_{\psi' 
inel}\rho(b,z)\right)$ gives the probability that the produced $\psi'$ has 
no inelastic collision in the nucleus, i.e.\ that it survives on the way 
out of the nucleus.

Then, in the semiclassical Glauber-approximation the cross section to 
subsequently produce a $J/\psi$ in an antiproton-nucleus collision is
\begin{eqnarray}
& &\sigma\left(\bar p + A\to J/\psi+X \right)=\cr
& &2\pi\int db\cdot b\,dz_1\,dz_2\,\theta(z_2-z_1){n_p\over 
A}\rho(b,z_1)\sigma\left(\bar p + p\to 
\psi' \right)exp\left(-\int\limits_{-\infty}^{z_1} dz \sigma_{\bar p N 
inel}\rho(b,z)\right)\cr
& &\times exp\left(-\int\limits_{z_1}^{z_2}dz\sigma_{\psi' N 
inel}\rho(b,z)\right)
 \sigma(\psi'+N\to\psi+N)\rho(b,z_2)
exp\left(-\int\limits_{z_2}^{\infty}dz\sigma_{\psi N 
inel}\rho(b,z)\right)\quad .\cr
& & 
\label{antipapsi}
\end{eqnarray}
The factor $exp\left(-\int\limits_{z_1}^{z_2}dz\sigma_{\psi' N 
inel}\rho(b,z)\right)$ 
is the probability to find, at the coordinate $(b,z_2)$, a $\psi'$ 
that was produced at $(b,z_1)$. $\sigma(\psi'+N\to\psi+N)\rho(b,z_2)$ is 
the 
probability that a $\psi'$ collides at the coordinate $(b,z_2)$ with a 
nucleon and that a $J/\psi$ is produced. Finally, 
$exp\left(-\int\limits_{z_2}^{\infty}dz\sigma_{\psi 
inel}\rho(b,z)\right)$ is the probability that the outgoing $J/\psi$ has 
no inelastic interactions. 
$\theta(z_2-z_1)=0$ for $z_2<z_1$, and $\theta(z_2-z_1)=1$ 
for $z_2>z_1$, this takes into account that a $\psi'$ has to be 
produced before it can collide again.

\begin{figure}
    \centering
        \leavevmode\
        \epsfxsize=.6\hsize
        \vspace{-1cm}
        \epsfbox{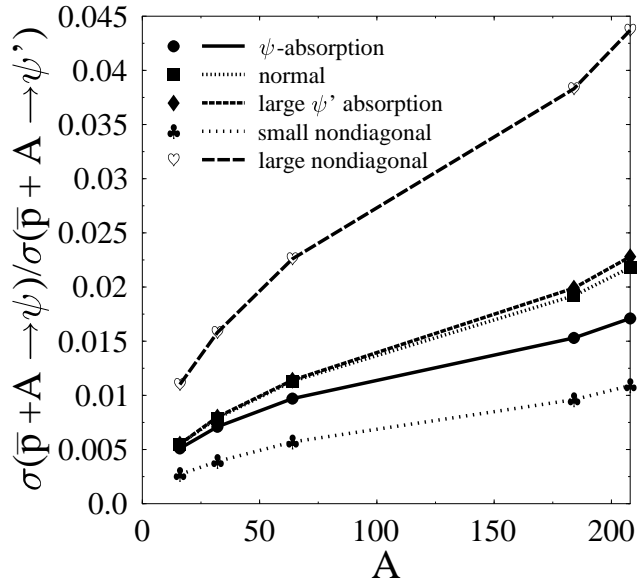}
\caption{The ratio $\sigma(\bar p +A \to 
\psi +nuclear fragments)/\sigma(\bar p +A \to 
\psi' +nuclear fragments)$ is shown for 5 different sets of 
parameters (see text for further details). Shown are the nuclear targets 
O, S, Cu, W, and Pb. The lines are just to guide the eye.
}
\label{antipa}
\end{figure}

In fig.~\ref{antipa} the ratio of the production cross section 
of the $J/\psi$ of eq.~(\ref{antipapsi}) to that of the $\psi'$
eq.~(\ref{antipapsiprim}) versus the size of the nuclear target
$A$ is plotted. Shown are the nuclear targets O, S, Cu, W, and Pb. The 
density distributions are from ref.~\cite{devries}. In contrast to 
eqs.~(\ref{antipapsiprim}) and~(\ref{antipapsi}), the ratio doesn't depend
on the cross section $\sigma\left(\bar p + p\to J/\psi+X \right)$, 
which is not well known at the threshold. 

In fig.~\ref{antipa} we used five sets of parameters. "normal" means
that the inelastic antiproton-nucleon cross section is $\sigma_{\bar p N
inel}=$50~mb, the inelastic cross section of the $\psi'$ is 
$\sigma_{\psi'N inel}=$7.5~mb, the inelastic cross section of the
$J/\psi$ is $\sigma_{\psi N inel}=$0~mb, and the cross section for the 
nondiagonal transition $\psi'+N\to J/\psi+N$ is
$\sigma_{\psi'+N\to\psi+N}=$0.2~mb. The other sets differ by only 
one of these parameters each:
\begin{itemize}
\item
In "$\psi$-absorption" $\sigma_{\psi N inel}=3.1$~mb.
\item
In "large $\psi'$ absorption" $\sigma_{\psi N inel}=15$~mb.
\item
In "small nondiagonal" $\sigma_{\psi' N \to \psi N}=0.1$~mb.
\item
In "large nondiagonal" $\sigma_{\psi' N \to \psi N}=0.4$~mb.
\end{itemize}
One can see that the result depends much more strongly on the nondiagonal 
cross section than on the absorption cross sections of the $J/\psi$ and 
the $\psi'$. Therefore, this process is well suited to measure the 
nondiagonal cross section. However, it will be necessary to differ 
between $J/\psi$'s produced in rescatterings of $\psi'$ and those which 
come from the decays of the $\psi'$
\footnote{The experiments also have to be able to detect  
radiative decays, because the $\psi'$ decays with a probability of 8.4\% 
and 6.4\% respectively into a photon and a $\chi_{c1}$ and a
$\chi_{c2}$ respectively, which decay with a probability 
of 31.6\% and 20.2\% respectively into a photon and a $J/\psi$~\cite{pdg}. 
In total this gives $8.4 \% \cdot 31.6 \% + 6.4 \% \cdot 20.2 \% = 3.9\%$, 
which is of the same order of magnitude as the nondiagonal transitions.}. 
However, $J/\psi$'s 
produced in $\psi'$ decays have longitudinal momenta which differ strongly 
from those produced in the two step processes and hence could be easily 
separated.

In antiproton-nucleus collisions at the $\psi'$ threshold, predominantly
$\psi'$ mesons are produced, i.e.\ the direct ($p\bar p \to J/\psi$)
production of $J/\psi$ mesons is strongly suppressed since this would
require huge Fermi momenta of the nucleons. Therefore, there are two
sources of $J/\psi$ production in this case. One source the nondiagonal
transitions where a produced $\psi'$ interacts in the final state via
$\psi'N \rightarrow J/\psi N$. The other source is the decay of $\psi'\to
J/\psi + X$, where $X$ could be two pions or other hadrons. Because the
$\psi'$s decay outside of the nucleus, the second reaction can be
eliminated experimentally. The momenta of the $J/\psi$s in these two
mechanisms are quite different. In particular, in the second mechanism
$y_{cms} \sim 0$ and the distribution is symmetric around $y=0$, while in
the nondiagonal mechanism there is a shift to rapidities closer to $y_A$.
For a rescattering without transfer of transverse momentum, this is a
shift of 0.2 units of rapidity, with transverse momentum transfer the
shift would be larger. In the center of mass system this is a shift of
$\Delta p_{cms}=0.6$~GeV. In the laboratory system a $J/\psi$ produced
at $y_{cms}=0$ has momentum of 6.1~GeV the $J/\psi$'s produced in
nondiagonal transitions have an average momentum of at most 4.2~GeV.

At the $J/\psi$ threshold the only possibility for an inelastic 
interaction of the produced $J/\psi$ is the channel 
$J/\psi+N\to\Lambda_c+D$. At the $\psi'$ threshold next to the 
channel $\psi'+N\to\Lambda_c+D$, $\psi'+N\to N +D+\bar D$ is 
also kinematically allowed. However, in the 
strange sector the channel $\phi+N\to N +K+\bar K$ is strongly 
suppressed versus the channel $\phi+N\to\Lambda+K$, therefore it 
is very likely that the $\Lambda_c+D$ channel dominates at both energies, 
the $J/\psi$ threshold and the $\psi'$ threshold.

\begin{figure}
    \centering
        \leavevmode\
        \epsfxsize=.6\hsize
        \vspace{-1cm}
        \epsfbox{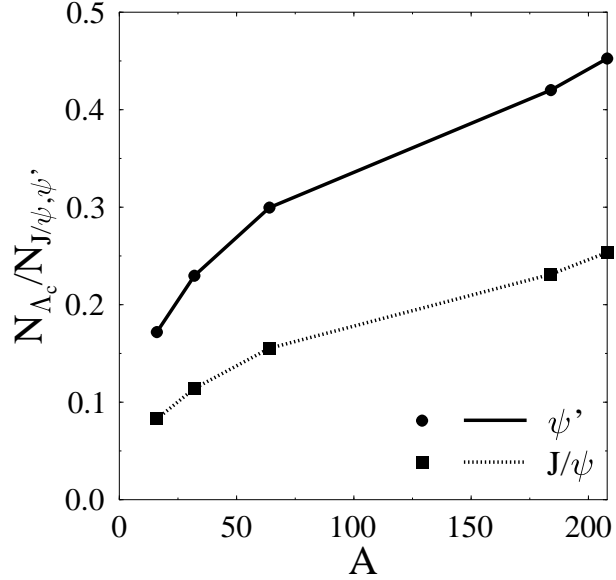}
\caption{The ratio of the number of $\Lambda_c$ divided by the number of 
produced $J/\psi$ and $\psi'$ respectively states at the threshold of 
$J/\psi$ and $\psi'$ production respectively. Shown are the nuclear 
targets O, S, Cu, W, and Pb. The lines are just to guide the eye.
}
\label{lambdac}
\end{figure}

The cross section for the production of $\psi'$ that 
doesn't undergo an inelastic rescattering is 
$\sigma\left(\bar p + A\to \psi' +\mbox{ nuclear 
fragments}\right)$ is given by eq.~(\ref{antipapsiprim}).
The cross section for the production of $\psi'$, whether they have 
subsequent inelastic scatterings or not is given by 
\begin{eqnarray}
& &\sigma\left(\bar p + A\to \psi' \right)_{w/o inel}=\cr
& &2\pi\int db\cdot b\,dz_1{n_p\over A}\rho(b,z_1)\sigma\left(\bar p +
p\to \psi' \right)exp\left(-\int\limits_{-\infty}^{z_1}
dz \sigma_{\bar p inel}\rho(b,z)\right)\quad.
\label{antipapsiprimwo}
\end{eqnarray}
Assuming that the $\Lambda_c$ channel is the only possible final state in 
inelastic collisions (i.e.\ the $D\bar D$ channel as well as the 
nondiagonal transition is neglected as a correction here), the fraction of 
the initially produced $\psi'$ that ends up in the $\Lambda_c$ channel is
\begin{equation}
{N_{\Lambda_c}\over N_{\psi'{initial}}}=1-{\sigma\left(\bar p + 
A\to \psi' +nuclear fragments 
\right)\over\sigma\left(\bar p + A\to \psi' +nuclear fragments 
\right)_{w/o inel}}\quad.
\label{lamdacfrac}
\end{equation}
Here we neglected the final state interactions of $\Lambda_c$ as they 
may only effect the momentum distribution of $\Lambda_c$ since the 
$\Lambda_c$ energy is below the threshold for the process 
$p+\Lambda_c \to N + N + D$. For this reaction the $\Lambda_c$ would need 
an energy of 4.2~GeV in the rest frame of the proton, while it has in 
average less than 3~GeV. 
The change of the momentum distribution of $\Lambda_c$ would provide 
unique information about the $\Lambda_c N$
interaction and could be a promising method for forming charmed 
hypernuclei.
Obviously eq.~\ref{lamdacfrac} is valid also for $\bar{D}$ production.
The fraction for the $\psi'$ and the $J/\psi$ 
threshold is depicted in 
fig.~\ref{lambdac}. Note that in eq.~(\ref{lamdacfrac}) and 
fig.~\ref{lambdac} the denominator is the number of produced particles, 
while in fig.~\ref{antipa} the denominator is the number of
$\psi'$ that don't undergo subsequent inelastic scatterings only. 

\section{Conclusion}

We found that accounting for color transparency phenomena within the
GVDM leads to a ratio of the elastic $J/\psi$ nucleon cross section to
the total $J/\psi$ nucleon cross section of approximately $5\%\div
6.5\%$. This value is significantly larger than that which follows from
the analysis based on VDM cf.~\cite{Ioffe:kk}, where $2\%$ was found.
Note that in ref.~\cite{Ioffe:kk} $|t_{eff}|=1/6\mbox{ GeV}^2$ has been
taken from soft QCD process of the photoproduction of $\rho$ mesons,
while the value following from the two gluon form factor extracted from
the photoproduction of $J/\psi$ mesons within the framework of the QCD
factorization theorem and the form factor of the $J/\psi$ is closer to
$|t_{eff}|=0.4\mbox{ GeV}^2$. Also, the total $J/\psi$-nucleon cross
section fitted with the VDM to the data is approximately a factor two
smaller than the value obtained within the GVDM.

This result is in agreement with naive expectations. The ratio of the
elastic cross section to the total cross section known for the pion or the
proton projectiles is approximately $25\%$. The total $J/\psi$-nucleon
cross section is significantly smaller than the total $\pi$-nucleon cross
section, because the $J/\psi$ has a much smaller size than the $\pi$ (the
radius of the $J/\psi$ is a factor 4 smaller the the one of the $\pi$).
Therefore the interaction of the $J/\psi$ with a nucleon is much weaker
than the $\pi$-nucleon interaction due to a stronger screening of color
charge of the constituents. That is also why the $J/\psi$~N interaction is
further from the black disk limit than the $\pi$. Measurements in $\bar p
A$ collisions will allow one to get new information about the strength of
the inelastic $J/\psi N$ and $\psi'N$ interactions leading to the
production of $\Lambda_c$ and $\bar{D}$ and for the first time about
nondiagonal transitions, $\psi' N\to J/\psi N$.

\vspace*{1cm}
{\bf Acknowledgement:}\\
We thank Ted Rogers for discussions.
This work was supported in part by BSF and DOE
grants. L.G.\ thanks the 
School of Physics and Astronomy of the Raymond and Beverly Sackler Faculty 
of Exact Science of the Tel Aviv University for support and hospitality.

\appendix

\section{Appendix: Amplitudes from the GVDM \label{app}} 

The imaginary part of the amplitude of the forward scattering of a 
$J/\psi$ on a 
nucleon is given by the 
cross section of eq.~(\ref{psiN}) and the optical theorem
\begin{equation}
I_{\psi\psi}=2\sqrt{s}p_{cm}\left(8\left({s\over 
s_{0}}\right)^{0.08}
+0.75\left({s\over s_{0}}\right)^{0.2}\right)\mbox{
GeV}^{-2}\quad.
\label{fpsiN}
\end{equation}
The real part of the amplitude can be evaluated with the help of 
the Gribov Migdal relation
\begin{equation}
R_{\psi\psi}={\sqrt{s}p_{cm}\pi\over 2}{{\partial \over \partial
\ln{s}}{I_{\psi\psi}\over \sqrt{s}p_{cm}}}\quad.
\label{gribovmigdal}
\end{equation}
This yields
\begin{eqnarray}
R_{\psi\psi}&=&{\pi}\sqrt{s}p_{cm}\left(0.64\left({s\over
s_{0}}\right)^{0.08}+0.15\left({s\over 
s_{0}}\right)^{0.2}\right)\mbox{ GeV}^{-2}
\quad.
\end{eqnarray}
The nondiagonal amplitudes and the amplitudes for the $\psi'$ is given in 
the GVDM by 
\begin{eqnarray}
f_{\gamma \psi}&=&{e\over f_{\psi}} f_{\psi \psi}+{e\over
f_{\psi'}}f_{\psi' \psi}\cr
f_{\gamma \psi'}&=&{e\over f_{\psi}}f_{\psi \psi'}
+{e\over f_{\psi'}} f_{\psi' \psi'}\quad,
\end{eqnarray}
$e$ is the charge of an electron and $f_{\psi}$ is the $J/\psi-\gamma$
coupling. $f_{\gamma \psi}$ is the $J/\psi$ photoproduction amplitude 
and $f_{\psi \psi}$ the 
$J/\psi$-nucleon elastic scattering amplitude.
The amplitudes $f_{\psi \psi'}$ and $f_{\psi' \psi}$ are the amplitudes 
for the nondiagonal transitions $J/\psi\to \psi'$ and $\psi'\to J/\psi$ 
respectively.

We assume that the real part of the amplitudes $f_{\gamma \psi}$ and 
$f_{\gamma \psi'}$ can be neglected. This is realistic, because 
$f_{\gamma \psi'}\ll f_{\psi \psi}$. 
Note that $f_{\psi\psi'}=f_{\psi' \psi}$, because of the CPT invariance 
of the amplitude. Then real parts of the two other amplitudes are 
\begin{eqnarray}
R_{\psi' \psi}&=&-{f_{\psi}\over f_{\psi'}}R_{\psi 
\psi}=-1.7 \cdot R_{\psi \psi}\cr
R_{\psi' \psi'}&=&-{f_{\psi}\over f_{\psi'}}R_{\psi'
\psi}=-1.7 \cdot R_{\psi' \psi}=2.9 \cdot R_{\psi \psi}
\quad.
\end{eqnarray} 
For their imaginary parts follows
\begin{eqnarray}
I_{\psi' \psi}&=&{f_{\psi'}\over e}I_{\gamma \psi}-{f_{\psi}\over 
f_{\psi'}}I_{\psi \psi}
=2\sqrt{s}p_{cm}\cdot 136.5 \cdot \left({s^{0.2}\over s_0^{1.2}}\right)
-1.7 \cdot I_{\psi \psi}
\cr
I_{\psi' \psi'}&=&{f_{\psi'}\over e}I_{\gamma \psi'}-{f_{\psi}\over 
f_{\psi'}}I_{\psi' \psi}=
{f_{\psi'}\over e}I_{\gamma \psi'}-1.7 \cdot I_{\psi' \psi}=
{f_{\psi'}\over e}I_{\gamma \psi'}+-1.7 \cdot {f_{\psi'}\over 
e}I_{\gamma \psi} +2.9 \cdot I_{\psi \psi}\cr
&=&2\sqrt{s}p_{cm}\cdot 175.5 \cdot \left({s^{0.2}\over s_0^{1.2}}\right)
+2.9 \cdot I_{\psi \psi}
\quad.
\end{eqnarray}

\end{document}